\documentclass[]{spie}

\usepackage{amsmath,amsfonts,amssymb}
\usepackage{graphicx}
\usepackage[colorlinks=true, allcolors=blue]{hyperref}
\usepackage{hyperref}
\usepackage{tikz}
\usepackage{orcidlink}

\newcommand{\orcid}[1]{\href{https://orcid.org/#1}{\textcolor[HTML]{A6CE39}{\aiOrcid}}}

\title{Construction of a Large Diameter Reflective Half-Wave Plate Modulator for Millimeter Wave Applications}

\author[a]{Joseph R.~Eimer \orcidlink{0000-0001-6976-180X}}
\author[a]{Michael K.~Brewer}
\author[b]{David T.~Chuss \orcidlink{0000-0003-0016-0533}}
\author[a]{John Karakla}
\author[a]{Rui Shi \orcidlink{0000-0001-7458-6946}}
\author[a]{John W.~Appel \orcidlink{0000-0002-8412-630X}}
\author[a]{Charles L.~Bennett \orcidlink{0000-0001-8839-7206}}
\author[a]{Joseph Cleary \orcidlink{0000-0002-7271-0525}}
\author[c]{Sumit Dahal \orcidlink{0000-0002-1708-5464}}
\author[a]{\mbox{Rahul Datta \orcidlink{0000-0003-3853-8757}}}
\author[c]{\mbox{Thomas Essinger-Hileman \orcidlink{0000-0002-4782-3851}}}
\author[a]{\mbox{Tobias A.~Marriage \orcidlink{0000-0003-4496-6520}}}
\author[a]{\mbox{Carolina N\'{u}\~{n}ez \orcidlink{0000-0002-5247-2523}}}
\author[f]{\mbox{Matthew~A.~Petroff \orcidlink{0000-0002-4436-4215}}}
\author[e]{\mbox{Duncan J. Watts \orcidlink{0000-0002-5437-6121}}}
\author[c]{\mbox{Edward J. Wollack \orcidlink{0000-0002-7567-4451}}}
\author[d]{Zhilei Xu \orcidlink{0000-0001-5112-2567}}

\affil[a]{The William H. Miller III Department of Physics and Astronomy, Johns Hopkins University, 3701 San Martin Drive, Baltimore, MD
21218, USA}
\affil[b]{Department of Physics, Villanova University, 800 Lancaster Avenue, Villanova, PA 19085, USA}
\affil[c]{NASA Goddard Space Flight Center, 8800 Greenbelt Rd, Greenbelt, MD 20771, USA}
\affil[d]{MIT Kavli Institute, Massachusetts Institute of Technology, 77 Massachusetts Avenue, Cambridge, MA 02139, USA}
\affil[e]{Institute of Theoretical Astrophysics, University of Oslo, P.O. Box 1029 Blindern, N-0315 Oslo, Norway}
\affil[f]{Center for Astrophysics, Harvard \& Smithsonian, 60 Garden Street, Cambridge, MA 02138, USA}

\authorinfo{Further author information: (Send correspondence to J. Eimer)\\J. Eimer: E-mail: jeimer1@jhu.edu}

\begin{document}
\maketitle

\begin{abstract}
Polarization modulation is a powerful technique to increase the stability of measurements by enabling the distinction of a polarized signal from dominant slow system drifts and unpolarized foregrounds. Furthermore, when placed as close to the sky as possible, modulation can reduce systematic errors from instrument polarization. 
In this work, we introduce the design and preliminary drive system laboratory performance of a new 60 cm diameter reflective half-wave plate (RHWP) polarization modulator. 
The wave plate consists of a wire array situated in front of a flat mirror. 
Using \mbox{50 $\mu$m} diameter wires with \mbox{175 $\mu$m} spacing, the wave plate will be suitable for operation in the millimeter wavelength range with flatness of the wires and parallelism to the mirror held to a small fraction of a wavelength. 
The presented design targets the 77--108 GHz range. Modulation is performed by a rotation of the wave plate with a custom rotary drive utilizing an actively controlled servo motor.
\end{abstract}

\section{Introduction}
The richness of many astrophysical processes ranging from supernova remnants and diffuse galactic magnetic fields to the cosmic microwave background (CMB) can be revealed through careful measurement of polarized emission \cite{Tinbergen96}. 
In fact, it is often the distribution of polarization over an extended source that contains the physical information of interest. For such observations, a measurement platform must provide a sufficiently stable calibration for the consistent interpretation of the signal over the angular and temporal scales of interest. 
In practice, these measurements can require high accuracy, the emission may be only slightly polarized, and instrument effect may contaminate the measurement; these challenges can be addressed through the use of modulation. 

In the millimeter and submillimeter wavelength range, polarization modulation can be achieved by placing a device in the optical path capable of varying the incoming polarized light in a known way while leaving the overall total intensity unchanged. 
Examples of such devices include half-wave plates (HWP)\cite{Jones88, Platt91, Leach91, Johnson07} and variable-delay polarization modulators (VPMs) \cite{Krejny08, Chuss2012, Harrington2018}. 
When combined with polarization sensitive detectors, such an optical system is capable of distinguishing between potentially larger unpolarized continuum flux and the targeted polarized signal. 

Polarization modulation has proven critical for far infrared and millimeter polarimetry, which includes the study of magnetic field structure, synchrotron emission, polarized thermal dust emission, and the CMB \cite{Krejny08, Wiesemeyer14, Johnson07, Takakura17, Harrington16, Kusaka14}.  
In this work, we present the design for a new reflective half-wave plate (RHWP) modulator of sufficient size to be suitable for use as a front-end polarization modulator for study of the CMB polarization at large angular scales. The design breaks new ground in terms of diameter, flatness, and system control\cite{Williams10, Wiesemeyer14}. 
Many design choices, e.g., the wave plate diameter, the modulation frequency, and the targeted wavelength are easily adopted to many different systems. 
As a specific goal, the design is optimized to fit in the 90 GHz telescope of the Cosmology Large Angular Scale Surveyor (CLASS) \cite{Harrington2018}. 

While the device is of broad general interest, using the 90 GHz CLASS telescope for an initial demonstration has many attractive features. 
For example, the telescope is already designed for a reflective front-end modulator such that no additional relay optics are required for integration, the calibration and optical performance of the CLASS telescopes are well understood and therefore commissioning can specifically target the RHWP, and using CLASS will enable a unique and important comparison of systematic and efficiency performance metrics between the CLASS VPM and this new RHWP modulator.

The paper is organized as follows. Section \ref{modulator_model} introduces a simple model describing how the RHWP can be used as a modulator. Section \ref{mechanical} describes the design of the wave plate and its expected mechanical performance. 
Section \ref{drive} describes how the modulator motion is controlled and gives preliminary performance from laboratory measurements. 
Finally we provide conclusions in Section \ref{conclusion}. 

\section{Operation of a reflective wave plate modulator}\label{modulator_model}

As a motivation for this instrument, a simplified model that demonstrates the concept of how the RHWP operates is presented. 
A more complete description can be found in the literature\cite{Wiesemeyer14}, and a more complete electromagnetic model will be included in future work. 

To describe the optical operation of the RHWP, we use Mueller calculus for the description of polarized light \cite{Shurcliff66}. 
The objective of a half-wave plate is to effectively introduce half a wavelength of path length difference between orthogonal linear polarization states. 
Regardless of the physical system used to implement such a transformation, the effect in the Stokes basis $(I, Q, U, V)$ can be represented via the Mueller matrix
\[
M_{WP} = \left(\begin{array}{cccc}
1 & 0 & 0 & 0 \\
0 & 1 & 0 & 0 \\
0 & 0 & -1 & 0 \\
0 & 0 & 0 & -1
\end{array}\right),
\]
where the chosen Stokes basis is oriented such that the wave plate ``fast'' axis is aligned with that of the $Q$ basis. In this orientation, the half-wave delay is introduced between the linear polarization states that define $Q$ such that the sign of $U$ is flipped; thus, the $-1$ in the $UU$ element of $M_{WP}$. Rotation of the wave plate in a fixed lab coordinate system can then be expressed as
\begin{equation}
\label{mueller_wp}
    M_{WP}(\theta) = R(-\theta) \cdot M_{WP} \cdot R(\theta) =
\left(\begin{array}{cccc}
1 & 0 & 0 & 0 \\
0 & \cos 4 \theta & \sin 4 \theta & 0 \\
0 & \sin 4 \theta & -\cos 4 \theta & 0 \\
0 & 0 & 0 & -1
\end{array}\right),
\end{equation}
where $R(\theta)$ is the Stokes basis rotation matrix:

\[
R(\theta) = \left(\begin{array}{cccc}
1 & 0 & 0 & 0 \\
0 & \cos (2 \theta) & \sin (2 \theta) & 0 \\
0 & -\sin (2 \theta) & \cos (2 \theta) & 0 \\
0 & 0 & 0 & 1
\end{array}\right).
\]
When followed by a detector sensitive to linear polarization, the polarization signal is then modulated at \emph{four} times the rotation rate of the RHWP.
To operate as a modulator, $\theta$ can be varied in a known way for subsequent analysis of the polarized component. 
A good review that places the half-wave plate modulator in context with alternative modulation strategies can be found in Figure 1 of Chuss et al.~(2012)\cite{Chuss12b}.

\begin{figure}[t]
    \centering
    \includegraphics[width=4in]{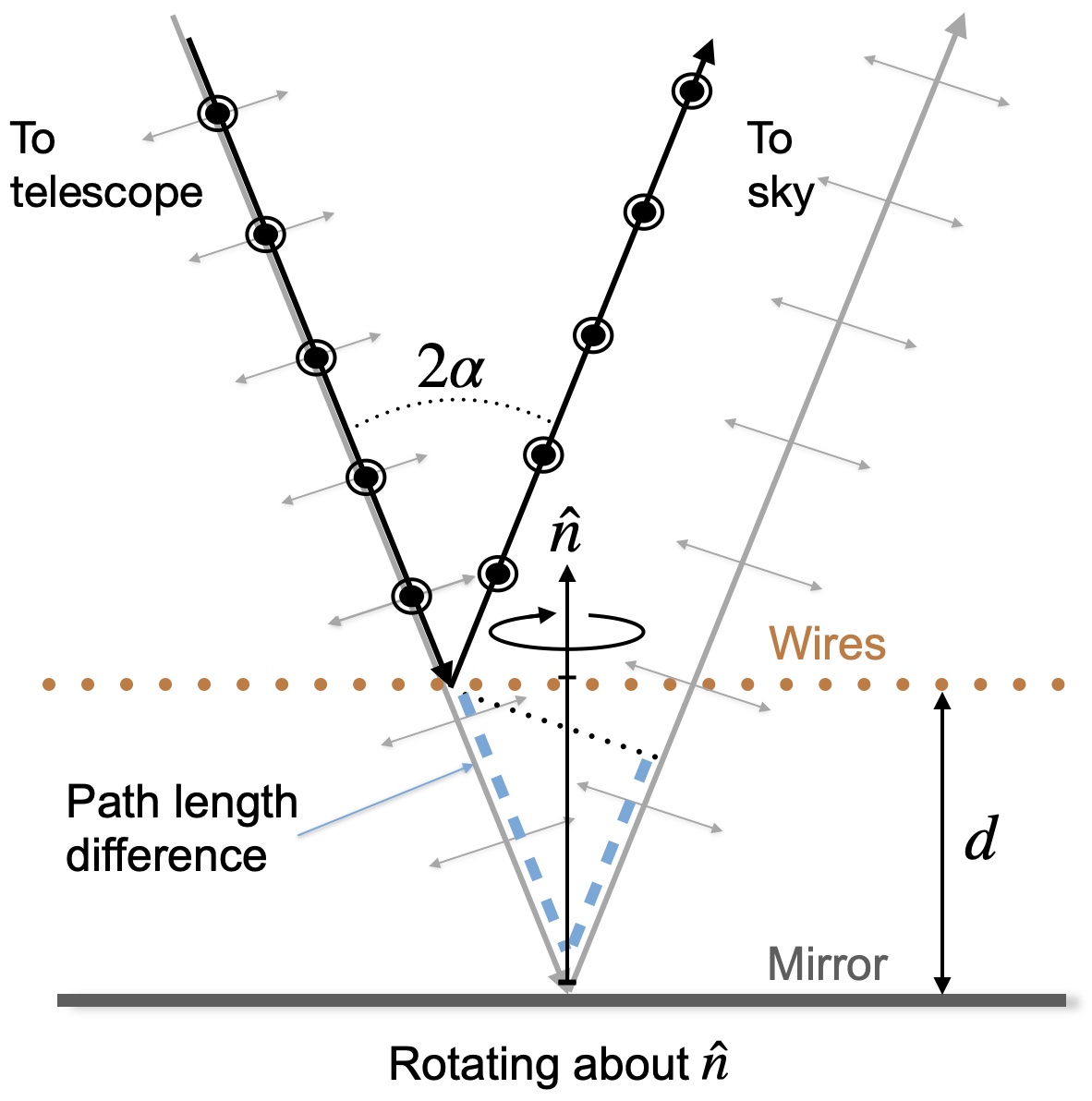}    \caption{Schematic of a reflective wave plate. The wire--mirror distance, $d$, is fixed to incorporate a half-wave path length difference between signals polarized parallel to the wires and those orthogonal to the wires. Modulation is achieved through rotation of the full wire and mirror assembly.}
    \label{fig:wp_cartoon}
\end{figure}

To achieve this effect in a reflective wave plate, note that the incoming and outgoing signals are easily distinguished by operating at nonzero incident angle, $\alpha$. 
In this configuration, the polarization state of an incoming signal can be split with a reflective wire array polarization filter. Light with polarization along the wires is reflected while the orthogonal state is transmitted; see Figure \ref{fig:wp_cartoon}. 
In the limit where the wires are much smaller than the wavelength, the relative phase between the two reflected states is given by 
\begin{equation}
\label{eq:phase}
    \phi = \frac{4 \pi d}{\lambda} \cos{\alpha},
\end{equation}
where $d$ is the distance between the wires and the mirror and $\lambda$ is the wavelength of the signal. 
The distance $d$ and the angle $\alpha$ are chosen so that $\phi = \pi$, i.e., the signal reflected by the mirror has traveled an additional half a wavelength. 
In this design, the angle of incidence is set to $\alpha=22.2^\circ$, and the nominal central wavelength is $\lambda=3.26$ mm to be consistent with the CLASS 90 GHz telescope \cite{Harrington2018}.
Following Equation \ref{eq:phase}, the wire--mirror spacing is set to be $d=0.88$ mm. 
In practice, observations occur over some finite bandwidth, and the total impact of modulation is the integrated effective delay experienced by each frequency in that band \cite{Wiesemeyer14}.

\section{Wave Plate Mechanical Design}\label{mechanical}

The architecture for this RHWP is flexible and has the ability of being tailored for use in a broad range of frequencies and optical systems. From Equation \ref{eq:phase}, one needs only adjust wire--mirror distance to operate in a different frequency band. 
Packaging and illumination profile concerns can also be addressed by using a diameter for the modulator that is suitable for the particular optical system. 
Since this particular design is optimized for testing on the CLASS 90 GHz telescope, the diameter is set to match the current modulator for that telescope, 60 cm.
This choice ensures consistent edge illumination to the operating telescope optics and avoids concerns from interference in packaging. 

\begin{figure}
    \centering
    \includegraphics[width=\linewidth]{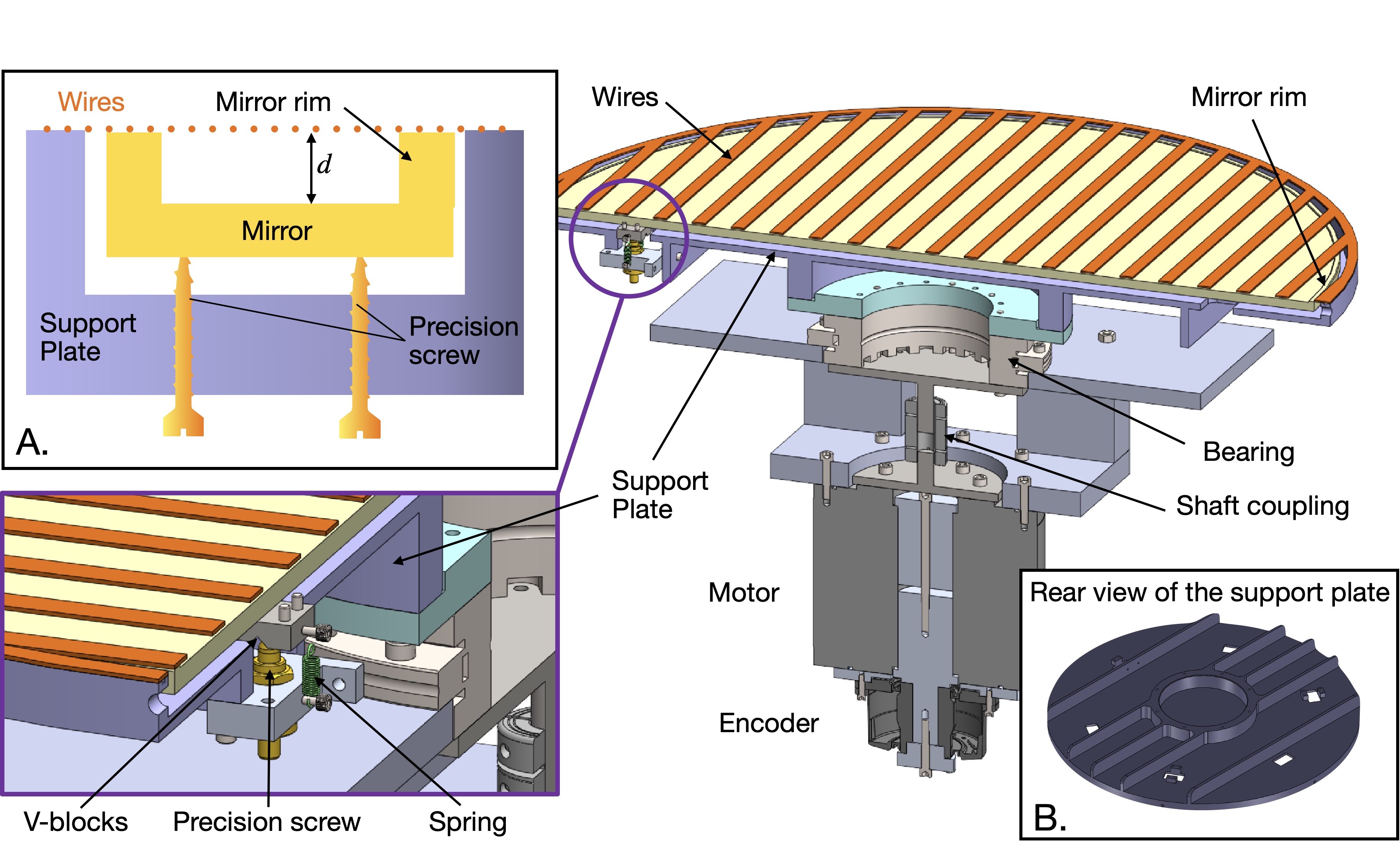}
    \caption{{\it Box A.} A cross section schematic for the RHWP concept. 
    Wires are bonded along the perimeter of the support plate. Precision screws can position the mirror with a raised rim into contact with the wire array. {\it Box B.} Rear view of the support plate. 
    The support plate uses five stiffening ribs running parallel to the wire direction to support the bending moment from the wire tension on the plate's opposite side. The plate is mounted to the rotation drive through the central ring. 
    {\it Background.} Cross section view of the model of the rotating half-wave plate system. 
    The wires are schematically illustrated with the sheet of thin bars over the mirror. The actual wires would not be visible in this scale. The full wave plate assembly is mounted on an interface bearing. The opposite side of the bearing holds the drive motor and absolute rotary encoder.}
    \label{fig:assem}
\end{figure}

Broadly speaking, the wave plate consists of three components: a wire array, a stiff support plate, and a flat mirror; see Figure \ref{fig:assem}, Box A. Part of the innovative structure of this design is that the stiff support structure is decoupled from the precision flat requirements of the mirror. 
This relaxes the stiffness requirements of the wire support and allows the entire assembly to be lighter weight and more modular.

The stiff support plate holds a uniformly-spaced wire array. 
Between the wires and support plate, a mirror is held with 127 threads per inch fine adjustment screws -- AJS127-0.5H. 
The mirror is augmented with a raised rim around its perimeter with height equal to the desired wire--mirror separation $d$. 
Full contact between the mirror rim and wires is the only requirement for specification of the phase delay. 
This means that the mirror rim also functions as a metrology flat for defining the plane of the wires. Once in contact, the surface normal vectors for the wire array and the mirror are both parallel, and alignment with the axis of rotation of the RHWP is achieved through fine adjustment of the mirror. 
By design, the contact between the wires and the mirror ensures the wires are moved in concert with the mirror during such an adjustment. 


The full RHWP modulator system concept is shown in Figure \ref{fig:assem}. 
In addition to the wave plate itself, a rotary drive system suitable for stable motion control and position registration is included. 
The drive is a critical aspect of the full modulator; the wave plate itself is insufficient. 
Details of the drive will be discussed further in \mbox{Section \ref{drive}.} 
In the following subsections the construction of the wire array, \mbox{\ref{sub_wires}}, the targeted wire tension, \mbox{\ref{sub_wireT}}, the design and performance of the support plate, \mbox{\ref{sub_plate}}, and the design and performance of the mirror, \mbox{\ref{sub_mirror}}, are described.

\subsection{Wire grid fabrication}\label{sub_wires}

To make the wire array, fine-gauge wire is wrapped onto a cylinder following Novak et al.\cite{Novak89}. The circumference defines the length of the wires.  
After securing the wires to removable bars running the length of the cylinder, the wires are cut between the bars, and the wire sheets are unwrapped from the cylinder and transferred to an intermediate stretching frame. Once brought to the desired tension, the wires are transferred to their final support structure. 
This precision wire grid polarizer fabrication procedure builds upon the work of others\cite{Payne78, Costley77, Lahtinen99} and has been extended to larger grids\cite{Voellmer2008}. 
The PIPER balloon-borne CMB experiment significantly improved the large mandrel wrapping technique\cite{Chuss2014}, which was further adapted for the CLASS VPMs\cite{Harrington2018}. 

Following the same reasoning as used for previous large grids\cite{Voellmer2008}, 50 $\mu$m diameter tungsten wire is used for the array. 
For increased electrical conductivity, the tungsten is plated with a $\sim 2.5$ $\mu$m thick layer of copper, using a nickel adhesion layer. 
Assuming a bulk resistivity of $1.72 \times 10^8$  $\Omega\cdot$m at 300 K, the penetration scale of the field into the wire plating is $\approx0.21\mu$m at a wavelength of 3 mm. 
Thus the wire plating is expected to be sufficiently thick to determine the emission properties of the wire. 

A wire spacing of 175 $\mu$m is targeted. 
This wire diameter to spacing ratio strikes a reasonable balance between the inductance for the polarization that is parallel to the wires and the capacitance of the perpendicular polarization. 
This broadly optimizes the performance of the wire grid as a quasioptical polarization diplexer\cite{Chuss2012}.

We briefly summarize the wrapping procedure in this paper; a more complete description has been described in previous literature\cite{Harrington2018}. 
An 8-inch diameter mandrel is used into which two bars are sunk.  These bars protrude slightly above the surface of the mandrel and provide locations for the eventual attachment points of each wire. The first step in the wrapping process is to cut grooves in these mandrel bars that will register each wire with the proper spacing relative to its neighbors. 
These grooves are cut into the bars using an end mill set to cut at a 45$^\circ$ angle. In this configuration, as the mandrel is turned, the end mill cuts `V'-shaped grooves in the mandrel bars at the desired wire spacing. This process takes place on a customized CNC milling machine, and the rotation of the axis is synced to the axial motion of the mandrel to set this spacing.

\begin{figure}[t]
    \centering
    \includegraphics[width=\textwidth]{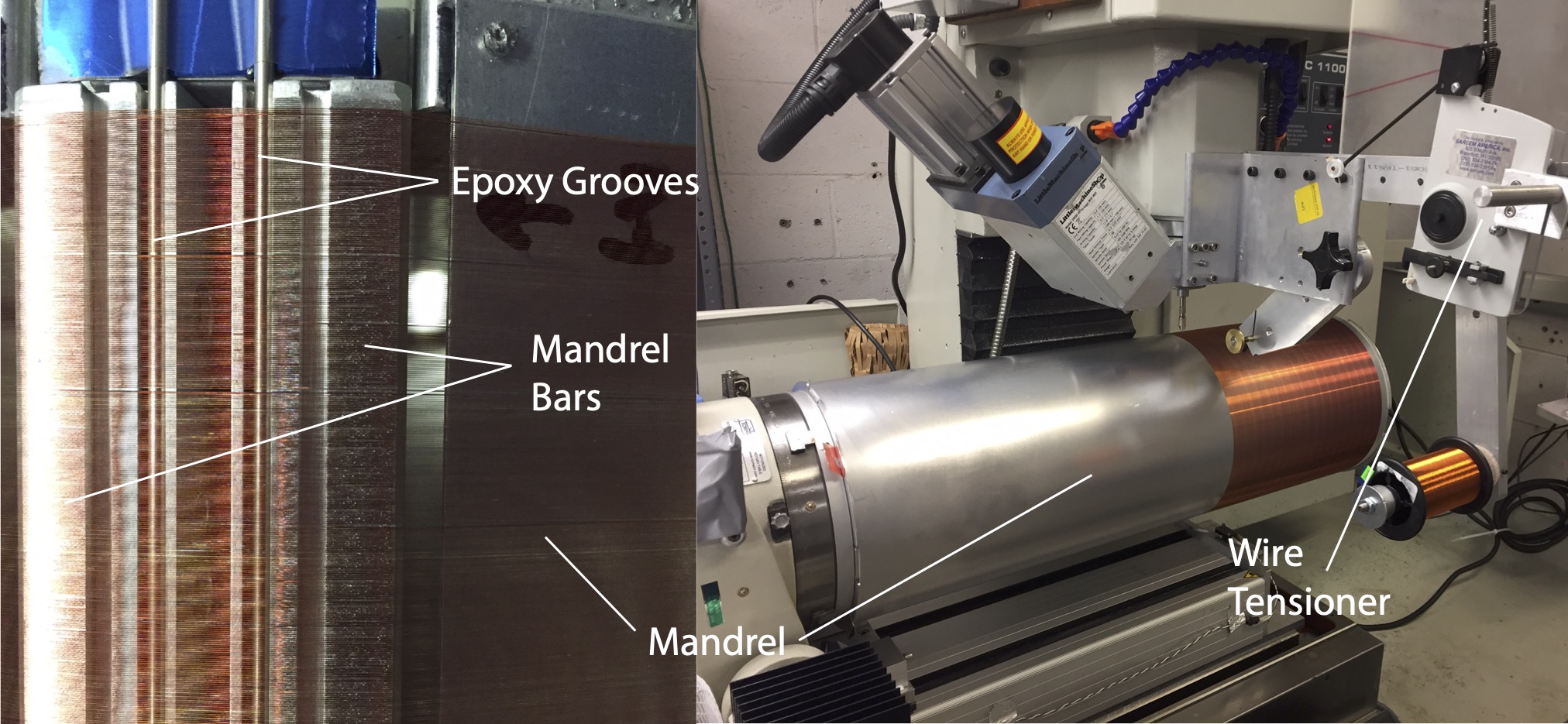}
    \caption{{\it Left.} A close up image of the wires after being wrapped on the mandrel. The two mandrel bars are seen running vertically, and the wires are crossing horizontally. {\it Right.} The custom setup for wrapping the wires on the mandrel is shown. The mandrel is mounted on an auxiliary rotary stage. As the stage rotates, the wire is drawn from a supply spool onto the mandrel.}
    \label{fig:gridwrap}
\end{figure}

After cutting the grooves in the bars, the wire is wrapped around the mandrel. The wire is fed from its spool through a wire tensioner to maintain approximately constant tension during the wrapping process and avoid sudden tension spikes that might break the wire. The synchronized motion of the CNC is utilized here to ensure that the wires are wrapped to lay in the grooves produced by the end mill. Images of the mandrel bars and wrapping process are shown in Figure~\ref{fig:gridwrap}.
After wrapping is complete, Stycast 2850FT epoxy (with Catalyst 23LV) is used to secure the wires to the mandrel bars.  The wires are then cut between the mandrel bars, and the wires are ready to be unwrapped.

Once unwrapped from the mandrel, the bars supporting the wires are transferred to a stiff frame capable of loading the wires with the desired level of tension \cite{Voellmer2008}. 
Rather than mounting the mandrel bars against flat registration pulling bars in the tensioning frame, the frame includes curved bars against which the mandrel bars are pulled. 
In this way, the relative tension of the wires across the array is tailored to meet the targets described in Section \ref{sub_wireT} below. The wires are then transferred to the support frame, described in Section \ref{sub_plate}. 

\subsection{Wire tension}\label{sub_wireT}

Previous measurements of the sky from Cerro Toco suggest that a polarization modulation frequency of \mbox{10 Hz} is an adequate target to elevate the polarization signal above the $1/f$ knee of low frequency noise \cite{Harrington21, Kusaka14}. 
To avoid excitation of vibration in the wires, which would degrade the uniformity of the phase delay definition, the fundamental frequency, $f$, of the wires is required to be much higher than $10~\mathrm{Hz}$. 

For this wire grid, an innovative strategy is adopted of varying the wire tension in accordance with the length of the wire in a way that the resonance frequency remains roughly constant across the face of the wave plate. 
Using this strategy, the overall tension on the ring supporting the wires is considerably less than if the wire tension was kept constant across the array. 
Under less tension, the support structure requires less mass to limit deformation to a given level. 

\begin{figure}[t]
    \centering
    \includegraphics[width=0.48\linewidth]{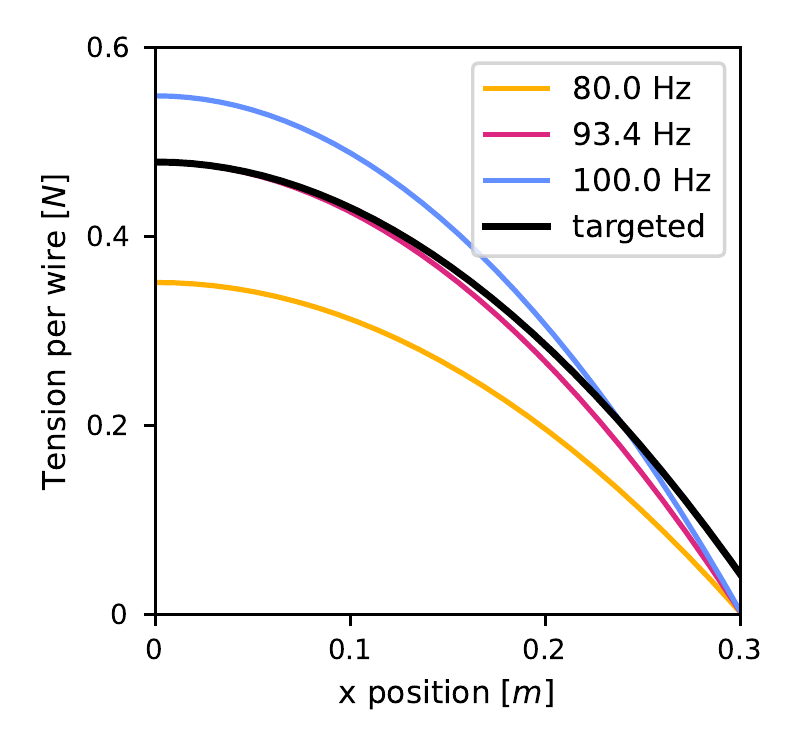}
    \caption{Wire tension as a function of its $x$ position, for different fundamental frequencies. The black thick line is the target initial tension, which includes a $980~\mathrm N$ component that is proportional to $(R_0^2-x^2)$, and a $20~\mathrm N$ component that is proportional to $x^2$. The smallest fundamental frequency on the targeted curve is $93.4~\mathrm Hz$.}
    \label{fig:wireTx}
\end{figure}

The tension of each wire is a function of its position in the array. Assuming a fixed resonance frequency $f$, the tension is given as
\begin{equation}\label{eq:wire_tension}
\begin{aligned}
T(x)&=4f^2m(x)L_0(x)\\
&=16f^2\rho\pi r^2(R_0^2-x^2)
\end{aligned}
\end{equation}
where $L_0(x)=2\sqrt{R_0^2-x^2}$ and $m(x)=\rho\pi r^2L_0(x)$ are the length and mass of the wire as a function of its $x$ position; directions are defined by the coordinate system shown in Figure \ref{fig:platedeform}. $R_0=30~\mathrm{cm}$ is the radius of the mirror; $\rho=19.3~\mathrm{g/cm^3}$ is the density of tungsten, and $r=25~\mathrm{\mu m}$ is the radius of the wire. Figure \ref{fig:wireTx} shows the wire tension $T(x)$ for different fundamental frequencies.

Based on a suite of simulations to predict the overall deflection of the support plate, see Section \ref{sub_plate}, a total load from the wires on the support plate of $1000~\mathrm{N}$ was found to have a good balance between a high wire resonance frequency and a reasonably light-weight support plate. 
The form of the total tension includes two parts: a $980~\mathrm{N}$ component that is proportional to $R_0^2-x^2$, cf. Equation \ref{eq:wire_tension}, and an added $20~\mathrm{N}$ component proportional to $x^2$ to prevent the tension from going to $0$ at the edges of the mirror. 
Without the offset, wires close to the edges could become loose after the support plate deforms under the tension from the wires.
The black curve in \mbox{Figure \ref{fig:wireTx}} illustrates the targeted initial tension of the wires across the array while being held with the tensioning frame.

\begin{figure}[t]
    \centering
    \includegraphics[width=0.8\linewidth]{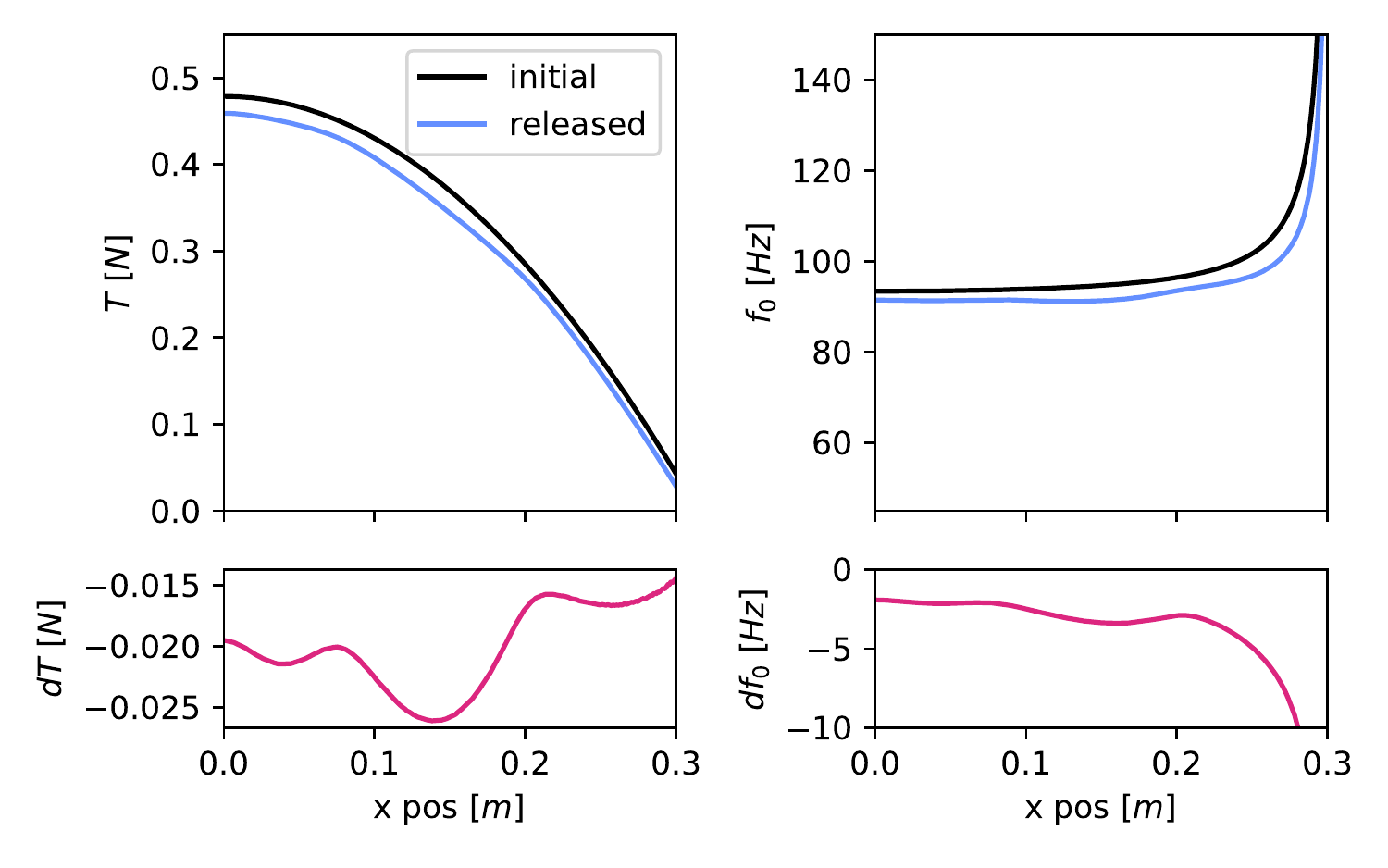}
    \caption{\textit{Upper left}: the initial (black) and released (blue) wire tension. 
\textit{Upper right}: The fundamental frequency corresponding to the initial (black) and released (blue) wire tension. The lower panels are showing the difference between initial and released quantities. The smallest resonance frequency on the released curve is $91.5~\mathrm{Hz}$. The small variations in the difference result from the positions of the stiffening ribs backing the support plate.}
    \label{fig:wiredTdf}
\end{figure}

The initial tension profile from the tensioning frame is slightly relaxed once the wires are bonded to the support plate perimeter and the load on the tensioning frame is released. 
The differences in tension, and resulting changes in resonance frequency, are sufficiently small that pre-loading the support plate can be omitted. Incorporating this tension change after release, the resonance frequency of the wires is found to be $ >91.5~\mathrm{Hz}$ across the entire array, as seen in Figure \ref{fig:wiredTdf}.

\subsection{Design of the support plate}\label{sub_plate}

The support plate is designed to hold the wires of the RHWP. While the final position of the wires is determined by the raised rim of the mirror, the support plate is designed to limit the need of the mirror flattening correction.
From this motivation, five ribs are added along the side of the plate opposite the wires. 
By aligning the ribs to support against the bending moment from the wires, the plate deformation is reduced without adding much mass; see Figure \ref{fig:assem}, Box B. 
The number of ribs, their locations, and their heights were chosen to minimize the mass of the plate while keeping the deformation in the $z$-direction less than $\pm 100~\mathrm{\mu m}$. 
This limit is sufficiently small to ensure the flattening ring can further reduce residual deformations to a small fraction of a wavelength. 
From this initial result, small adjustments were made to improve manufacturability of the design, add margin to the total deformation, and properly interface with other parts of the RHWP system. 
These modifications reduce the deflection from the initial 200 $\mu$m range to only $\sim 34$ $\mu$m at the expense of a modest increase in mass. 
For a different application requiring even less deformation or mass, the ribs could be further optimized by making them taller, different heights, or configured with T-shaped cross sections. 

\begin{figure}[t]
    \centering
    \includegraphics[width=0.48\linewidth]{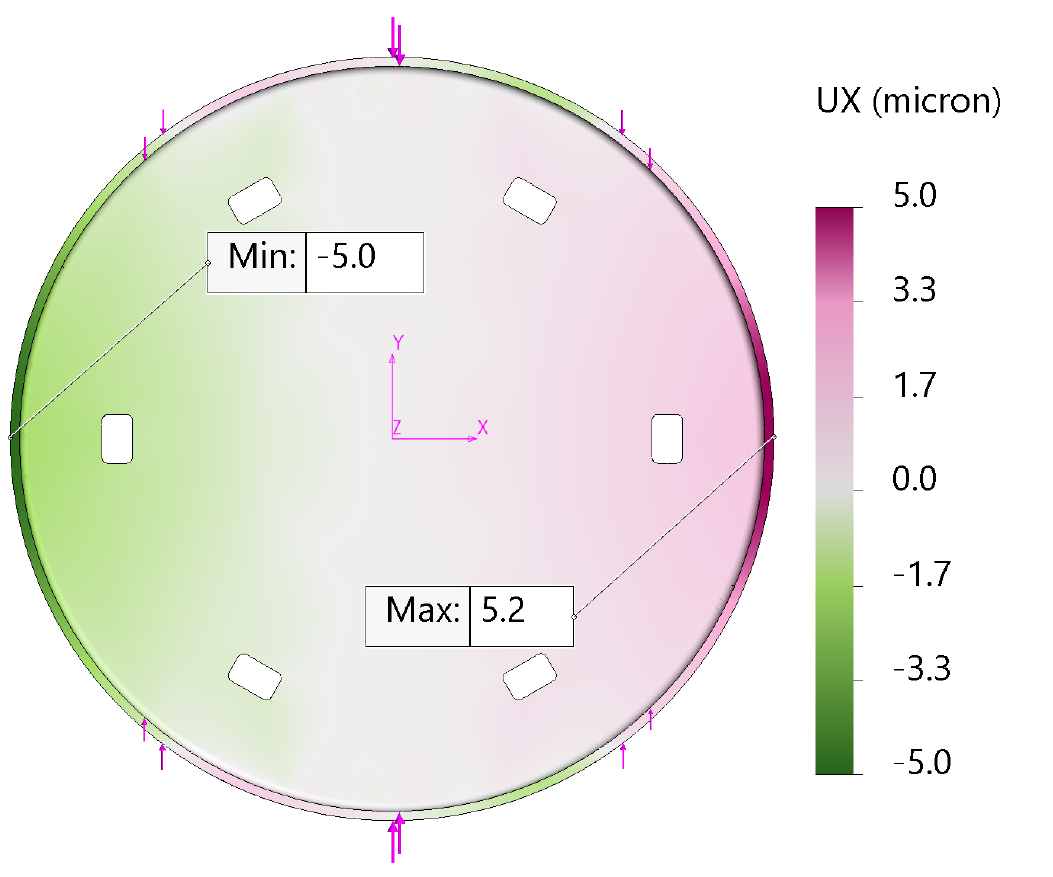}
    \includegraphics[width=0.48\linewidth]{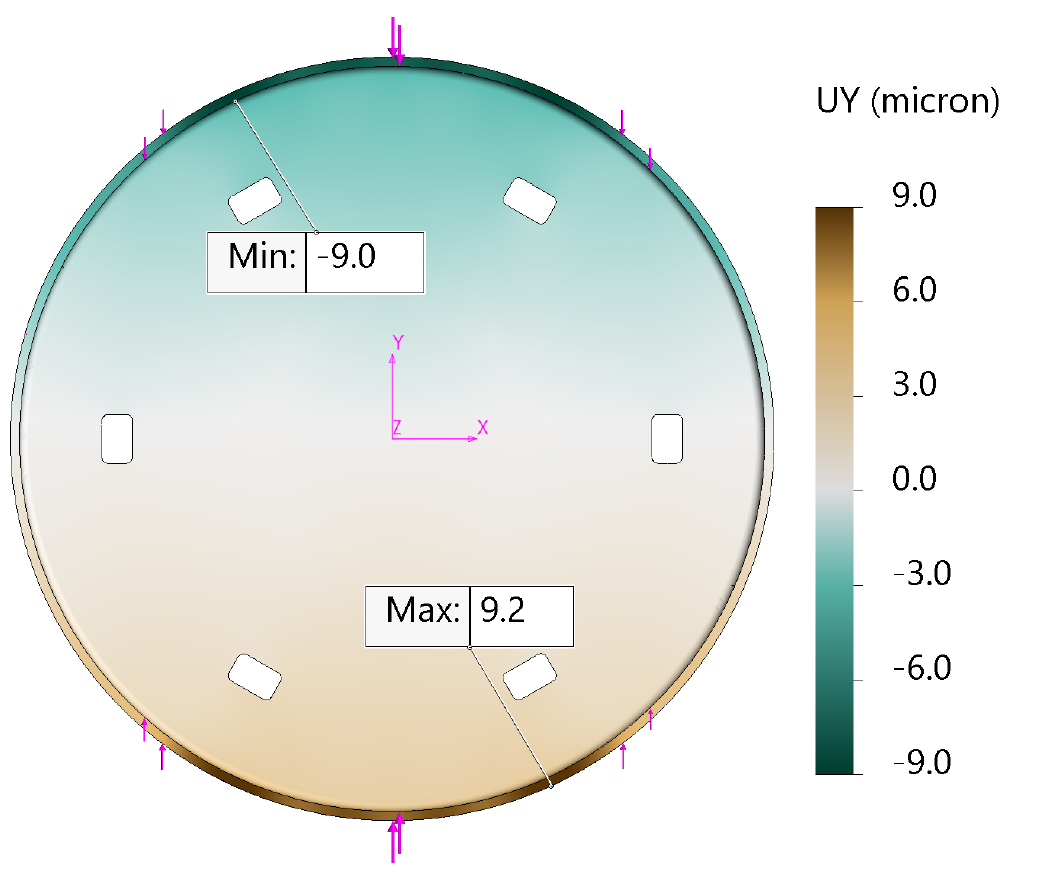}
    ~\\~\\
    \includegraphics[width=0.48\linewidth]{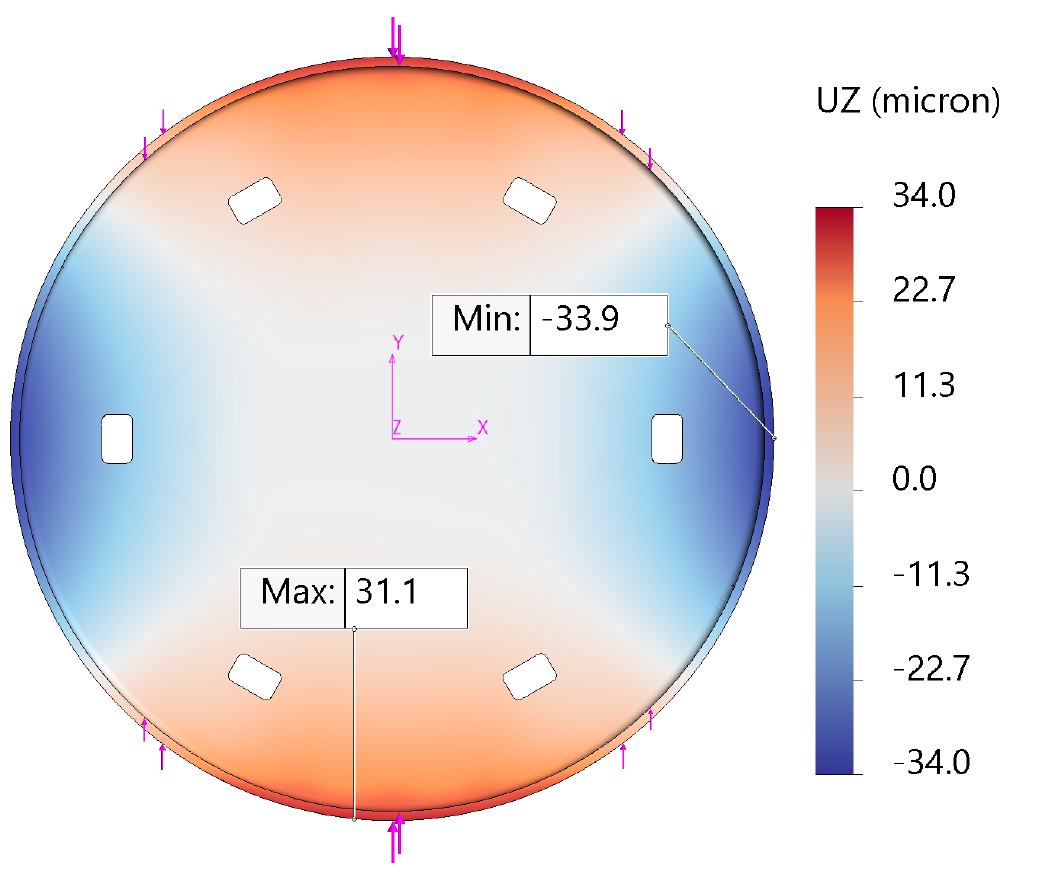}
    \includegraphics[width=0.48\linewidth]{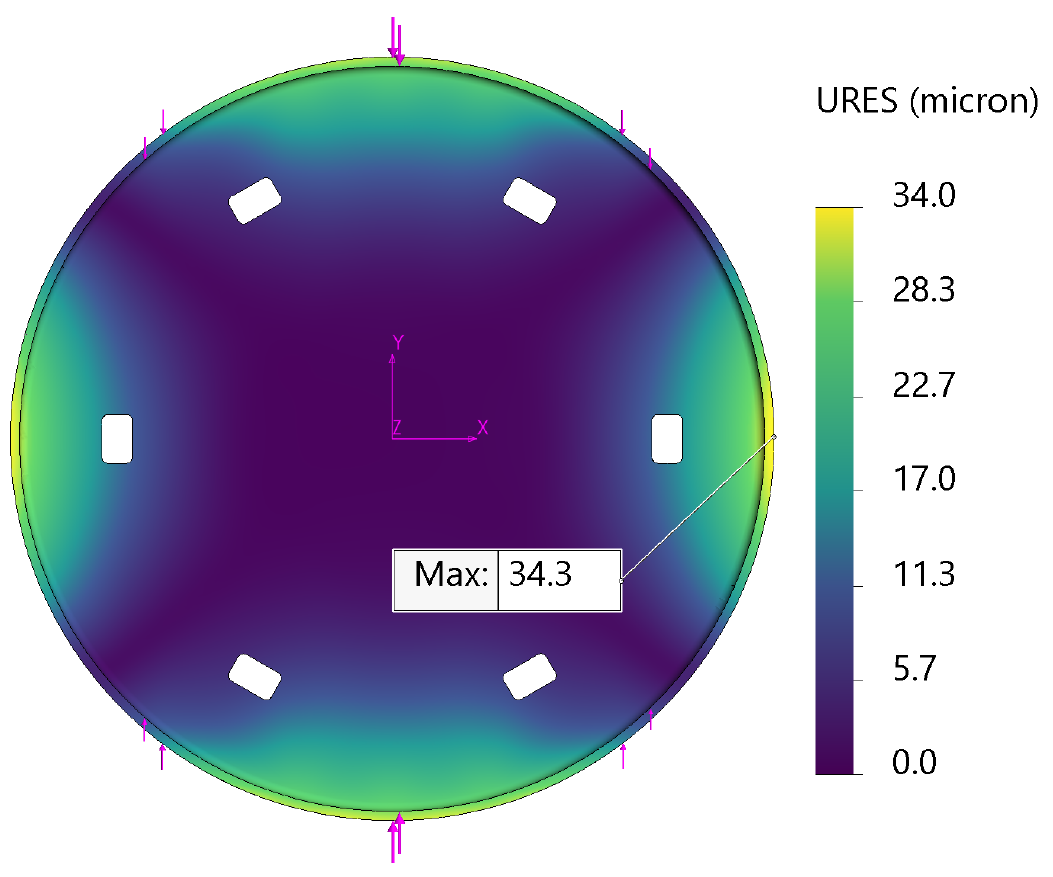}
    \caption{The FEA simulated plate displacements along $x$-axis, labeled UX (\textit{top-left}), $y$-axis, labeled UY (\textit{top-right}), $z$-axis, labeled UZ (\textit{bottom-left}), and the total amplitude, labeled URES (\textit{bottom-right}).
    The wire initial tension is applied to the top surface of the plate edge ring, and its amplitude and direction are represented by the arrows arround the circumference of the plate in each panel.
    The total force applied on the upper/lower half-ring is $1000~\mathrm{N}$.
    The coordinate system is shown in the center of each panel, in which the $z$-axis is pointing towards the reader.
    Most of the deformation comes from the $z$-axis displacements with a range of $65~\mathrm{\mu m}$, which is comparable to the wire diameter.}
    \label{fig:platedeform}
\end{figure}

Figure \ref{fig:platedeform} shows the finite element analysis (FEA) simulation results of the support plate deformation from the initial tension of the wires. 
A non-uniformly-distributed $1000~\mathrm{N}$ tension, see Figure \ref{fig:wireTx}, is applied on the surface of the support plate perimeter rim.
The $\mathrm{UY}$ plot, upper right, indicates that the tension on the wires will decrease after the plate deforms, which implies these results represent the upper limit on the deformation amplitudes. 

Using $\mathrm{UX}$, upper left, the impact on the wire spacing of the array is estimated. 
The wire pitch is found to change by less than $+17.6~\mathrm{nm}$, close to the edge, and $-8.5~\mathrm{nm}$, close to the center; this is a negligible fraction of the ($175~\mathrm{\mu m}$) target pitch.

The change in wire tension and resonance frequency $f$ is estimated by considering $\mathrm{UY}$; the resulting upper limits of the final tension and resonance frequency are shown with the blue curves in the upper panels of \mbox{Figure \ref{fig:wiredTdf}}. The smallest $f$ after wires are released is $91.5~\mathrm{Hz}$ --  much higher than the $10~\mathrm{Hz}$ modulation frequency.

\subsection{Design of the flattening ring mirror}\label{sub_mirror}

As described above, the mirror piece integrates the flat mirror with a delay defining raised rim. 
Furthermore, the entire mirror structure is designed to be stiff enough to resist the force from pressing against the wires and drawing them into a flat configuration. 
This section reviews the results of FEA simulations of the final shape of the mirror while in contact with the wires. 

\begin{figure}[t]
    \centering
    \includegraphics[width=0.4\linewidth]{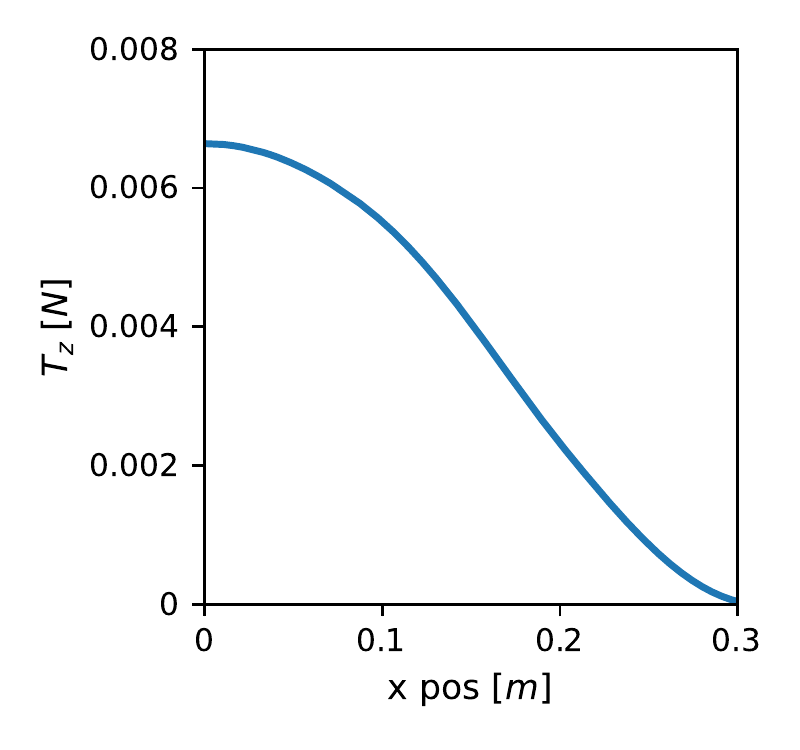}\quad
    \includegraphics[width=0.55\linewidth]{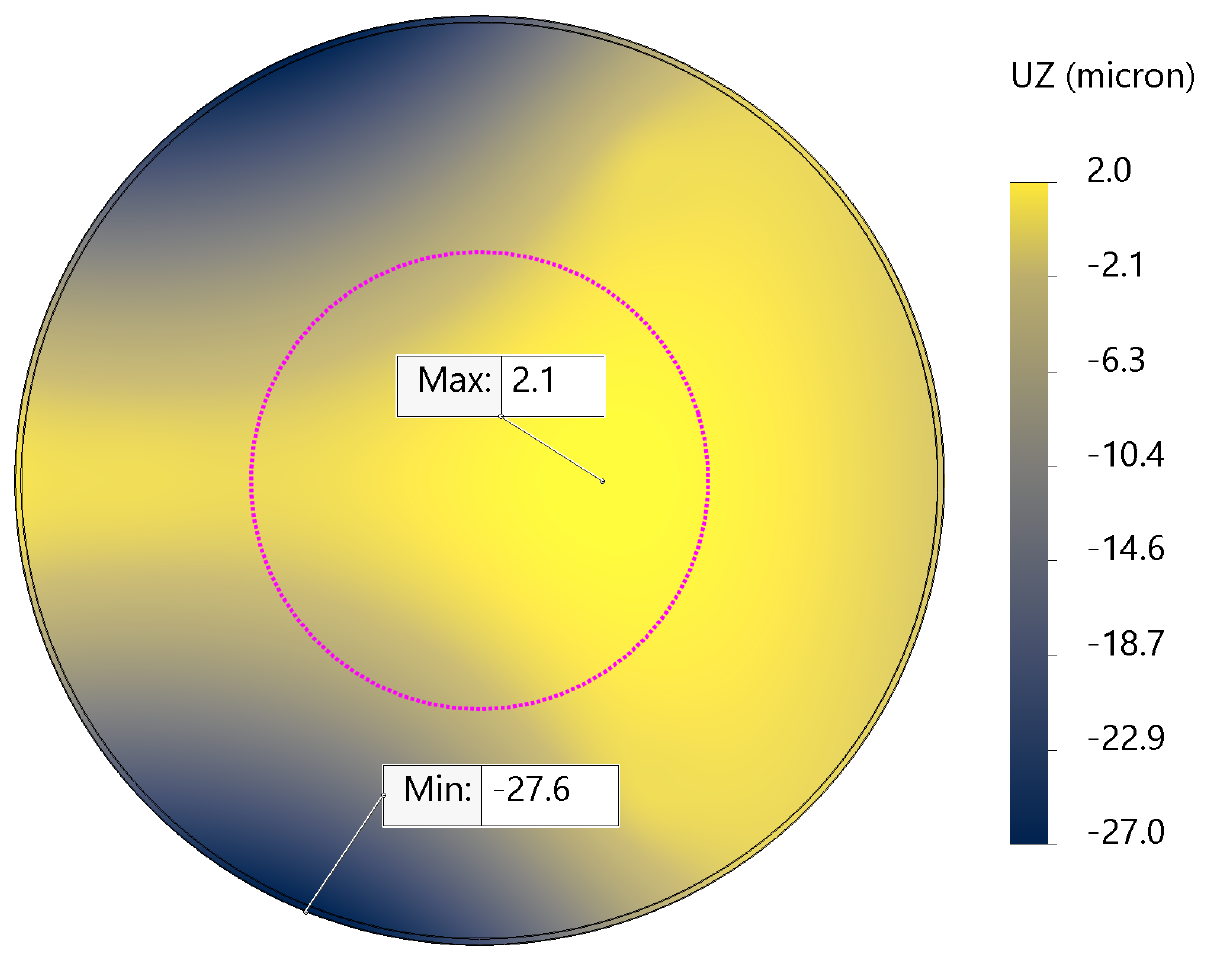}
    \caption{\textit{Left}: The $z$-axis projection of the wire tension on the mirror flattening ring.
    \textit{Right}: The FEA simulated $z$-axis deformation of the mirror.
    A total of $22.5~\mathrm{N}$ force that obeys the distribution shown in the left panel is applied on the mirror flattening ring.
    The $z$-axis displacement range across the whole mirror is $29.7~\mathrm{\mu m}$, but the range within the illuminated region (inside magenta circle) is from $-5.3~\mathrm{\mu m}$ to $2.1~\mathrm{\mu m}$. The visible trefoil deformation pattern results from the selected location of the three mounting points.}
    \label{fig:mirrordeform}
\end{figure}

The force on the mirror perimeter is determined by using the $\mathrm{UZ}$ deformation results; see Figure \ref{fig:platedeform} lower left. 
When pushing the mirror towards the wires, the flattening ring will first touch the wires closer to the left and right edge. 
To make the wires flat, the mirror must move forward for at least the range of the $z$-direction deformation.
The mirror will then support the projection of wire tension along the $z$-axis.
When computing the projected force on the mirror rim, in addition to the $\mathrm{UZ}$ result, a uniformly distributed $25~\mathrm{\mu m}$ displacement is added as an operational buffer.
The projected force on the face of the plate is shown in the left panel of \mbox{Figure \ref{fig:mirrordeform}}.
This corresponds to a total of $22.5~\mathrm{N}$ force normal to the top surface of the mirror rim.
The simulated mirror deformation along the $z$-axis is shown in the right panel of Figure \ref{fig:mirrordeform}.
The displacement range across the whole mirror is $29.7~\mathrm{\mu m}$. 
This result assumes an \emph{overestimated} deformation of the support plate, since the assumed wire load was from the initial wire tension and not from the released tension. 
The support plate is therefore expected to deform \emph{less} than shown in Figure \ref{fig:platedeform}, and the mirror will be required to support \emph{less} load along its rim. 
Given this estimation method, this final flatness prediction represents a conservative estimate of the mirror deformation. At this level, the final surface is more likely to be dominated by machining tolerances, expected to be $< \pm 20 \mu$m. 
Depending on the application, the mirror plate could be made with a larger overall thickness -- potentially with lightweighting features -- to achieve a flatness level consistent with the project objectives. 
For initial testing on the CLASS telescope, the edges are under-illuminated; a 30 cm diameter circle centered on the RHWP is a reasonable estimate of the illuminated region \cite{Eimer12}. In this region, see the magenta circle in Figure \ref{fig:mirrordeform}, the displacement range is from $-5.3~\mathrm{\mu m}$ to $2.1~\mathrm{\mu m}$.

As mentioned above, the mirror is held using three high precision screws.
Each screw contacts a V-groove along a block mounted to the rear of the mirror. 
The three grooves are aligned with the center of the mirror. 
The V-blocks and the tip of the adjustment screws are of similar hardness stainless steel alloys to avoid potential binding or pitting between these mating surfaces.  
The mirror is held fast against the V-blocks by two extension springs neighboring each adjustment screw.
With this method of mounting, the external stress from the support plate and any thermo-mechanical stress from temperature changes in the operating environment cannot be transferred to the mirror itself.

\section{RHWP Drive}\label{drive}

Our initial operational plan for the RHWP modulator is to execute rotation using a single direction, constant angular velocity motion. 
In reality, the wave plate will be mounted in a telescope system executing scanning maneuvers and being buffeted by winds. 
In order to isolate the rotation from these external factors, a servo drive system with sufficient bandwidth and control loop speed to produce stable output motion has been developed. 
The drive system, including position readout, is a critical component for the success of the modulator. Without precision phase control and accurate position knowledge, demodulation of the data products can become cumbersome at best and possibly contaminated by significant systematic errors and instabilities. 
Therefore the drive system is a \emph{critical} part of the modulator system itself, and the current configuration for our drive is described here. One interesting feature of the reflective wave plate approach is that the drive system naturally lives behind the wave plate mirror. Not only does this help separate drive electronics from the optical path, reducing risk from radio frequency interference, but it also enables the use of conventional bearings, motors, and encoders for essentially arbitrary diameter devices.

Due to the requirement for both velocity and position control, a programmable servo drive and a brushless AC servo motor are used to position the RHWP. 
The motor is a 16 pole direct-drive servo motor from Kollmorgen, DH062M-12-1310. 
This motor is able to control a load with several hundred times the moment of inertia of the armature  without the need for gear reduction, which provides the advantage of being able to incorporate the motor, load, and position readout encoder all on one shaft. 
The servo drive is an ABB/Baldor MicroFlex e100,  
and the controlling computer is a PICMG 1.3 single board computer running the VxWorks real time operating system mounted in a mixed PCI/PCI Express passive backplane.

The servo loop runs at 20 Hz, taking as input the commanded position and velocity one time step in the future along with the current commanded acceleration. 
When running at a constant velocity, the commanded position is simply incremented by the velocity times the time step, and the commanded acceleration is zero. 
The loop first queries the servo drive for its current motor encoder position and velocity along with various other quantities such as current, voltage, following error, etc. 
This query uses the Baldor Immediate Command Mode protocol (ICM), which provides the ability to include these queries into a signal TCP packet, and receive the data packed into one packet as well. 
Typical turn around time is about 2.2 ms over 100 Mbps Ethernet -- more than adequate for the 50 ms servo loop time step. 
The loop then calls a profiler, which uses these quantities to calculate the required velocity and acceleration for the next time step. 
These are then packed into an ICM packet and sent to the drive. 

The position readout encoder is a Heidenhain RCN 2581 angle encoder with a system accuracy of $\pm 2$ seconds of arc. 
This encoder uses an Endat serial data interface with a resolution of 28 bits per revolution for the absolute position readout. 
It also has 1 volt peak-to-peak analog sine/cosine incremental signals with a resolution of 16384 cycles per revolution. 
The absolute position is read out by a PCI Express card on the control computer. 
The readout is done at $\sim 201$ Hz and is activated by an interrupt sent by the same clock that triggers the detector data reads -- ensuring synchronous recording for later demodulation. 
The interrupt handler first latches the time on a PCI GPS card also installed in the control computer. 
It then latches the encoder signal and passes off to a lower priority interrupt service routine, which reads the clock and encoder, differences the encoder position to derive velocity, gathers all of the additional data read from the servo drive at 20 Hz, and then packs all of this into a data structure written to a ring buffer. 
A third task running at 1 Hz empties the ring buffer and then writes the data to disk.

Our current test setup uses the full system shown in Figure \ref{fig:assem}, except the wave plate is replaced by a metal disk replicating the expected 0.627 Nm moment of inertia of the RHWP. To minimize risk from pulse width modulation motor drive currents, a three phase electromagnetic interference filter with $\sim 250$ kHz cutoff is added. 
While the motor has an internal Heidenhain ECN 113 rotary encoder, we use the sine/cosine outputs from the higher precision RCN 2581 for motor position feedback. 
This increases the feedback resolution by a factor of 8 and significantly reduces velocity error due to the much better linearity of the higher precision encoder.
Laboratory testing in front of a cold receiver with transition edge sensor bolometers using a similar setup showed no discernible interference in the relatively high noise environment of the room.
Velocity control of this preliminary lab test is demonstrated to 0.006\% at the 2.5 Hz rotational velocity of the RHWP. 


\section{Conclusion}\label{conclusion}
Front-end polarization modulation is a powerful tool for distinguishing a faint polarized signal from a much larger unpolarized background while protecting against systematic errors such as instrument polarization. 
In this work, the design of a new \mbox{60 cm} clear aperture reflective wave plate has been demonstrated. 
The architecture is flexible and can be scaled for a wide range of applications. 
In the prototype, the parallelism of polarizer and mirror are anticipated to meet our need through the constraints in the design. 
Similarly, the deformation of the mirror in its final configuration is estimated to be limited to less than \mbox{29 $\mu$m.}

A prototype drive system has been constructed in the lab using a realistic mass model, and early tests demonstrate the planned control system is feasible and enables a stable rotational frequency, normally \mbox{2.5 Hz}, with an acceptably small error. 

\acknowledgments

We acknowledge the National Science Foundation Division of Astronomical Sciences for their support. 
The RHWP modulator is being developed with support from Grant Numbers 2034400 and 2109311.  We further acknowledge the very generous support of Jim and Heather Murren (JHU A\&S ’88), Matthew Polk (JHU A\&S Physics BS ’71), David Nicholson, and Michael Bloomberg (JHU Engineering ’64). Zhilei Xu is supported by the Gordon and Betty Moore Foundation through grant GBMF5215 to the Massachusetts Institute of Technology.

\bibliography{references}

\begin{thebibliography}{10}

\bibitem{Tinbergen96}
J.~{Tinbergen}, {\em {Astronomical Polarimetry}}, 1996.

\bibitem{Jones88}
T.~J. {Jones} and D.~{Klebe}, ``{A Simple Infrared Polarimeter},'' {\em
  Publications of the Astronomical Society of the Pacific}~{\bf 100}, p.~1158,
  Sept. 1988.

\bibitem{Platt91}
S.~R. {Platt}, R.~H. {Hildebrand}, R.~J. {Pernic}, J.~A. {Davidson}, and
  G.~{Novak}, ``{100-micron Array Polarimetry From the Kuiper Airborne
  Observatory: Instrumentation, Techniques, and First Results},'' {\em
  Publications of the Astronomical Society of the Pacific}~{\bf 103}, p.~1193,
  Nov. 1991.

\bibitem{Leach91}
R.~W. {Leach}, D.~P. {Clemens}, B.~D. {Kane}, and R.~{Barvainis},
  ``{Polarimetric Mapping of Orion Using MILLIPOL: Magnetic Activity in
  BN/KL},'' {\em The Astrophysical Journal}~{\bf 370}, p.~257, Mar. 1991.

\bibitem{Johnson07}
B.~R. {Johnson}, J.~{Collins}, M.~E. {Abroe}, P.~A.~R. {Ade}, J.~{Bock},
  J.~{Borrill}, A.~{Boscaleri}, P.~{de Bernardis}, S.~{Hanany}, A.~H. {Jaffe},
  T.~{Jones}, A.~T. {Lee}, L.~{Levinson}, T.~{Matsumura}, B.~{Rabii},
  T.~{Renbarger}, P.~L. {Richards}, G.~F. {Smoot}, R.~{Stompor}, H.~T. {Tran},
  C.~D. {Winant}, J.~H.~P. {Wu}, and J.~{Zuntz}, ``{MAXIPOL: Cosmic Microwave
  Background Polarimetry Using a Rotating Half-Wave Plate},'' {\em The
  Astrophysical Journal}~{\bf 665}, pp.~42--54, Aug. 2007.

\bibitem{Krejny08}
M.~{Krejny}, D.~{Chuss}, C.~D. {D'Aubigny}, D.~{Golish}, M.~{Houde}, H.~{Hui},
  C.~{Kulesa}, R.~F. {Loewenstein}, S.~H. {Moseley}, G.~{Novak}, G.~{Voellmer},
  C.~{Walker}, and E.~{Wollack}, ``{The Hertz/VPM polarimeter: design and first
  light observations},'' {\em Applied Optics}~{\bf 47}, p.~4429, Aug. 2008.

\bibitem{Chuss2012}
D.~T. {Chuss}, E.~J. {Wollack}, R.~{Henry}, H.~{Hui}, A.~J. {Juarez},
  M.~{Krejny}, S.~H. {Moseley}, and G.~{Novak}, ``{Properties of a
  variable-delay polarization modulator},'' {\em Applied Optics}~{\bf 51},
  p.~197, Jan. 2012.

\bibitem{Harrington2018}
K.~{Harrington}, J.~{Eimer}, D.~T. {Chuss}, M.~{Petroff}, J.~{Cleary},
  M.~{DeGeorge}, T.~W. {Grunberg}, A.~{Ali}, J.~W. {Appel}, C.~L. {Bennett},
  M.~{Brewer}, R.~{Bustos}, M.~{Chan}, J.~{Couto}, S.~{Dahal}, K.~{Denis},
  R.~{D{\"u}nner}, T.~{Essinger-Hileman}, P.~{Fluxa}, M.~{Halpern},
  G.~{Hilton}, G.~F. {Hinshaw}, J.~{Hubmayr}, J.~{Iuliano}, J.~{Karakla},
  T.~{Marriage}, J.~{McMahon}, N.~J. {Miller}, C.~{Nu{\~n}ez}, I.~L. {Padilla},
  G.~{Palma}, L.~{Parker}, B.~{Pradenas Marquez}, R.~{Reeves}, C.~{Reintsema},
  K.~{Rostem}, D.~{Augusto Nunes Valle}, T.~{Van Engelhoven}, B.~{Wang},
  Q.~{Wang}, D.~{Watts}, J.~{Weiland}, E.~{Wollack}, Z.~{Xu}, Z.~{Yan}, and
  L.~{Zeng}, ``{Variable-delay polarization modulators for the CLASS
  telescopes},'' in {\em Millimeter, Submillimeter, and Far-Infrared Detectors
  and Instrumentation for Astronomy IX},  J.~{Zmuidzinas} and J.-R. {Gao},
  eds., {\em Society of Photo-Optical Instrumentation Engineers (SPIE)
  Conference Series} {\bf 10708}, p.~107082M, July 2018.

\bibitem{Wiesemeyer14}
H.~{Wiesemeyer}, T.~{Hezareh}, E.~{Kreysa}, A.~{Weiss}, R.~{G{\"u}sten}, K.~M.
  {Menten}, G.~{Siringo}, F.~{Schuller}, and A.~{Kovacs}, ``{Submillimeter
  Polarimetry with PolKa, a Reflection-Type Modulator for the APEX
  Telescope},'' {\em Publications of the Astronomical Society of the
  Pacific}~{\bf 126}, p.~1027, Nov. 2014.

\bibitem{Takakura17}
S.~{Takakura}, M.~{Aguilar}, Y.~{Akiba}, K.~{Arnold}, C.~{Baccigalupi},
  D.~{Barron}, S.~{Beckman}, D.~{Boettger}, J.~{Borrill}, S.~{Chapman},
  Y.~{Chinone}, A.~{Cukierman}, A.~{Ducout}, T.~{Elleflot}, J.~{Errard},
  G.~{Fabbian}, T.~{Fujino}, N.~{Galitzki}, N.~{Goeckner-Wald}, N.~W.
  {Halverson}, M.~{Hasegawa}, K.~{Hattori}, M.~{Hazumi}, C.~{Hill}, L.~{Howe},
  Y.~{Inoue}, A.~H. {Jaffe}, O.~{Jeong}, D.~{Kaneko}, N.~{Katayama},
  B.~{Keating}, R.~{Keskitalo}, T.~{Kisner}, N.~{Krachmalnicoff}, A.~{Kusaka},
  A.~T. {Lee}, D.~{Leon}, L.~{Lowry}, F.~{Matsuda}, T.~{Matsumura},
  M.~{Navaroli}, H.~{Nishino}, H.~{Paar}, J.~{Peloton}, D.~{Poletti},
  G.~{Puglisi}, C.~L. {Reichardt}, C.~{Ross}, P.~{Siritanasak}, A.~{Suzuki},
  O.~{Tajima}, S.~{Takatori}, and G.~{Teply}, ``{Performance of a continuously
  rotating half-wave plate on the POLARBEAR telescope},'' {\em Journal of
  Cosmology and Astroparticle Physics}~{\bf 2017}, p.~008, May 2017.

\bibitem{Harrington16}
K.~{Harrington}, T.~{Marriage}, A.~{Ali}, J.~W. {Appel}, C.~L. {Bennett},
  F.~{Boone}, M.~{Brewer}, M.~{Chan}, D.~T. {Chuss}, F.~{Colazo}, S.~{Dahal},
  K.~{Denis}, R.~{D{\"u}nner}, J.~{Eimer}, T.~{Essinger-Hileman}, P.~{Fluxa},
  M.~{Halpern}, G.~{Hilton}, G.~F. {Hinshaw}, J.~{Hubmayr}, J.~{Iuliano},
  J.~{Karakla}, J.~{McMahon}, N.~T. {Miller}, S.~H. {Moseley}, G.~{Palma},
  L.~{Parker}, M.~{Petroff}, B.~{Pradenas}, K.~{Rostem}, M.~{Sagliocca},
  D.~{Valle}, D.~{Watts}, E.~{Wollack}, Z.~{Xu}, and L.~{Zeng}, ``{The
  Cosmology Large Angular Scale Surveyor},'' in {\em Millimeter, Submillimeter,
  and Far-Infrared Detectors and Instrumentation for Astronomy VIII},  W.~S.
  {Holland} and J.~{Zmuidzinas}, eds., {\em Society of Photo-Optical
  Instrumentation Engineers (SPIE) Conference Series} {\bf 9914}, p.~99141K,
  July 2016.

\bibitem{Kusaka14}
A.~{Kusaka}, T.~{Essinger-Hileman}, J.~W. {Appel}, P.~{Gallardo}, K.~D.
  {Irwin}, N.~{Jarosik}, M.~R. {Nolta}, L.~A. {Page}, L.~P. {Parker},
  S.~{Raghunathan}, J.~L. {Sievers}, S.~M. {Simon}, S.~T. {Staggs}, and
  K.~{Visnjic}, ``{Publisher's Note: ``Modulation of cosmic microwave
  background polarization with a warm rapidly rotating half-wave plate on the
  Atacama B-Mode Search instrument'' [Rev. Sci. Instrum. 85, 024501 (2014)]},''
  {\em Review of Scientific Instruments}~{\bf 85}, p.~039901, Mar. 2014.

\bibitem{Williams10}
B.~D. Williams, {\em B-Machine Polarimeter: A Telescope to Measure the
  Polarization of the Cosmic Microwave Background}.
\newblock PhD thesis, University of California, Santa Barbara, 2010.

\bibitem{Shurcliff66}
W.~A. {Shurcliff}, {\em {Polarized light. Production and use}}, 1966.

\bibitem{Chuss12b}
D.~T. {Chuss}, E.~J. {Wollack}, G.~{Pisano}, S.~{Ackiss}, K.~{U-Yen}, and M.~w.
  {Ng}, ``{A translational polarization rotator},'' {\em Applied Optics}~{\bf
  51}, p.~6824, Oct. 2012.

\bibitem{Novak89}
G.~Novak, R.~J. Pernic, and J.~L. Sundwall, ``Far infrared polarizing grids for
  use at cryogenic temperatures,'' {\em Appl. Opt.}~{\bf 28}, pp.~3425--3427,
  Aug 1989.

\bibitem{Payne78}
J.~M. Payne and M.~R. Wordeman, ``Quasi‐optical diplexer for millimeter
  wavelengths,'' {\em Review of Scientific Instruments}~{\bf 49}(12),
  pp.~1741--1743, 1978.

\bibitem{Costley77}
A.~E. Costley, K.~H. Hursey, G.~F. Neill, and J.~M. Ward, ``Free-standing
  fine-wire grids: Their manufacture, performance, and use at millimeter and
  submillimeter wavelengths,'' {\em J. Opt. Soc. Am.}~{\bf 67}, pp.~979--981,
  Jul 1977.

\bibitem{Lahtinen99}
J.~Lahtinen and M.~Hallikainen, ``Fabrication and characterization of large
  free-standing polarizer grids for millimeter waves,'' {\em International
  Journal of Infrared and Millimeter Waves} , Jan 1999.

\bibitem{Voellmer2008}
G.~M. {Voellmer}, C.~{Bennett}, D.~T. {Chuss}, J.~{Eimer}, H.~{Hui}, S.~H.
  {Moseley}, G.~{Novak}, E.~J. {Wollack}, and L.~{Zeng}, ``{A large
  free-standing wire grid for microwave variable-delay polarization
  modulation},'' in {\em Ground-based and Airborne Instrumentation for
  Astronomy II},  I.~S. {McLean} and M.~M. {Casali}, eds., {\em Society of
  Photo-Optical Instrumentation Engineers (SPIE) Conference Series} {\bf 7014},
  p.~70142A, July 2008.

\bibitem{Chuss2014}
D.~T. {Chuss}, J.~R. {Eimer}, D.~J. {Fixsen}, J.~{Hinderks}, A.~J. {Kogut},
  J.~{Lazear}, P.~{Mirel}, E.~{Switzer}, G.~M. {Voellmer}, and E.~J. {Wollack},
  ``{Variable-delay polarization modulators for cryogenic millimeter-wave
  applications},'' {\em Review of Scientific Instruments}~{\bf 85}, p.~064501,
  June 2014.

\bibitem{Harrington21}
K.~{Harrington}, R.~{Datta}, K.~{Osumi}, A.~{Ali}, J.~W. {Appel}, C.~L.
  {Bennett}, M.~K. {Brewer}, R.~{Bustos}, M.~{Chan}, D.~T. {Chuss},
  J.~{Cleary}, J.~{Denes Couto}, S.~{Dahal}, R.~{D{\"u}nner}, J.~R. {Eimer},
  T.~{Essinger-Hileman}, J.~{Hubmayr}, F.~{Raul Espinoza Inostroza},
  J.~{Iuliano}, J.~{Karakla}, Y.~{Li}, T.~A. {Marriage}, N.~J. {Miller},
  C.~{N{\'u}{\~n}ez}, I.~L. {Padilla}, L.~{Parker}, M.~A. {Petroff},
  B.~{Pradenas M{\'a}rquez}, R.~{Reeves}, P.~{Flux{\'a} Rojas}, K.~{Rostem},
  D.~{Augusto Nunes Valle}, D.~J. {Watts}, J.~L. {Weiland}, E.~J. {Wollack},
  Z.~{Xu}, and {Class Collaboration}, ``{Two Year Cosmology Large Angular Scale
  Surveyor (CLASS) Observations: Long Timescale Stability Achieved with a
  Front-end Variable-delay Polarization Modulator at 40 GHz},'' {\em The
  Astrophysical Journal}~{\bf 922}, p.~212, Dec. 2021.

\bibitem{Eimer12}
J.~R. {Eimer}, C.~L. {Bennett}, D.~T. {Chuss}, T.~{Marriage}, E.~J. {Wollack},
  and L.~{Zeng}, ``{The cosmology large angular scale surveyor (CLASS): 40 GHz
  optical design},'' in {\em Millimeter, Submillimeter, and Far-Infrared
  Detectors and Instrumentation for Astronomy VI},  W.~S. {Holland} and
  J.~{Zmuidzinas}, eds., {\em Society of Photo-Optical Instrumentation
  Engineers (SPIE) Conference Series} {\bf 8452}, p.~845220, Sept. 2012.

\end{thebibliography}
\bibliographystyle{spiebib}
\end{document}